\documentclass[a4paper]{article}

\usepackage{epsfig}
\usepackage{graphicx}

\newcommand{\req}[1]{(\ref{#1})}
\newcommand{\nn}{\nonumber}

\def\muF{\mu^2_F}
\def\muO{\mu^2_0}
\def\eps{\epsilon}
\def\als{\alpha_s}
\def\qbq{q\overline{q}}
\def\ubu{{u}\overline{u}}
\def\dbd{{d}\overline{d}}
\def\sbs{{s}\overline{s}} 
\begin{document}

\begin{flushright}
IRB-TH-4/03
\end{flushright}

\vspace*{1.7cm}

\begin{center}

\begin{large}

{\bf 
     HARD EXCLUSIVE REACTIONS\\ AND THE TWO-GLUON COMPONENTS
     OF\\ $\eta$ AND $\eta'$ MESONS%
       \footnote{
       Presented at the
       Second International Conference on Nuclear and Particle physics 
       with CEBAF at Jefferson Laboratory (NAPP 2003), Dubrovnik, Croatia}
}

\end{large}

\end{center}

\begin{center}
        KORNELIJA PASSEK-KUMERI\v{C}KI\footnote{Electronic address: 
                             passek@thphys.irb.hr}
\end{center}

\begin{center}
\begin{small}
{\em Theoretical Physics Division, Rudjer Bo\v{s}kovi\'{c} Institute, 
        P.O. Box 180,\\ HR-10002 Zagreb, Croatia}
\end{small}
\end{center}

\begin{center}
\begin{small}
October 2003
\end{small}
\end{center}

\vspace*{0.35cm}
\noindent
The formalism for treating the leading-twist contributions
of the two-gluon Fock components occurring in hard exclusive processes
that involve $\eta$ and $\eta'$ mesons is reviewed.
The calculation
of the $\eta$, $\eta'$--photon 
transition form factor in next-to-leading order in $\als$,
as well as, 
the analysis of the $g^* g^* \eta'$ vertex
and the electro- and photoproduction of $\eta$, $\eta'$ mesons
are presented.
Applications of this formalism to other relevant quantities 
such as glueballs are also discussed.
\\

\vspace*{0.25cm}

\section{{\it{\Large Introduction}}}

Within the framework for analyzing exclusive processes 
at large momentum transfer developed in the late seventies\cite{sHSA},
the description of hard exclusive processes 
involving light mesons
is based 
on the factorization of short-
and long-distance dynamics 
and on the application of perturbative QCD. 
The former dynamics is represented by the
process-dependent and perturbatively calculable
parton-level subprocess amplitude, i.e.
elementary hard-scattering amplitude,
in which the meson is replaced by its Fock states,
while the latter is described by 
the process-independent meson distribution
amplitude (DA), 
which represents the probability of finding
the corresponding Fock state in a meson
and encodes the soft physics.
Although the DA 
is essentially a nonperturbative quantity, 
its evolution is subject to a perturbative treatment.
In the standard hard-scattering approach, the leading-twist
contributions are obtained by regarding the meson as consisting only
of valence Fock states, transverse parton momenta are neglected
(collinear approximation) as well as the masses.

This work is focused on hard reactions involving
$\eta$ and $\eta'$ mesons.
In the formalism explained above,  
these particles are naturally described in terms of
the  SU(3)$_F$ octet and singlet
valence quark-antiquark Fock components 
and the two-gluon component, which also carries 
the flavour-singlet quantum numbers. 
A separate distribution amplitude corresponds 
to each of the three components.
In comparison with the reactions involving the ``pure'' flavour-nonsinglet 
mesons ($\pi$, $K$, $\ldots$), the following novel features should be
properly taken into account.
First, 
owing to SU(3)$_F$ symmetry breaking and U(1)$_A$ anomaly, 
the well-known flavour mixing is present
in the $\eta$-$\eta'$ system 
(for a recent review, see \cite{Feldmann99}).
Second, there are three valent components that contribute 
to $\eta$ and $\eta'$ in leading-twist
and two of them are connected by evolution,
i.e. the mixing of the singlet and gluon DAs under
evolution should be properly taken into account. 
The latter feature has been investigated in a number 
of papers 
\cite{Terentev81,Ohrndorf81,ShifmanV81etc,BelitskyM98etc}.
However, although most of the results are in agreement
\cite{Terentev81,ShifmanV81etc,BelitskyM98etc} 
up to differences in the conventions used,
a consistent set of conventions necessary for the calculation
of both the elementary hard-scattering amplitude and the DA 
was not transparent from these works.
Recently, the treatment of the two-gluon component and its
mixing with the singlet one has been reexamined in \cite{KrollP02}.
A detailed analysis of the next-to-leading order (NLO) calculation
of the $\eta$, $\eta'$-photon transition form factor was performed,
making it possible to introduce and test the conventions
for all ingredients of a leading-twist calculation for any
process that involves $\eta$ or $\eta'$ mesons.
The results were then 
applied to the $\eta$, $\eta'$--gluon transition form factor and
the electroproduction of $\eta$ and $\eta'$ mesons.
In this work we briefly review the basic steps of that
analysis, stress the important points occasionally still 
overlooked in the literature and extend the application
of this formalism to photoproduction of $\eta$ and $\eta'$ mesons
and a possible description of glueballs.

\section{{\it{\Large Formalism}}}

As the valence Fock components of the pseudoscalar mesons
$P=\eta, \eta'$, we choose the SU(3)$_{\rm F}$ octet and singlet
combinations of quark-antiquark states 
\begin{equation}
\begin{array}{lcl}
|\qbq_{8}\rangle = |(\ubu+\dbd-2\sbs)/\sqrt{6} \rangle\,, 
& \: &
|\qbq_{1}\rangle = |(\ubu+\dbd+\sbs)/\sqrt{3} \rangle\,, 
\end{array}
\label{eq:osbasis}
\end{equation}
and the two-gluon state:
\begin{equation}
|gg\rangle \, .
\label{eq:ggstate}
\end{equation}
The corresponding DAs are denoted by $\Phi_{P8,1,g}$
and parameterized as
\begin{equation}
\begin{array}{lcl}
\displaystyle
\Phi_{P8}(x,\mu^2)=\frac{f_P^8}{2\sqrt{2 N_c}} \, \phi_8(x,\mu^2) 
\, ,
& \: &  \\[0.4cm]
\displaystyle
\Phi_{P1}(x,\mu^2)=\frac{f_P^1(\mu^2)}{2\sqrt{2 N_c}} \, \phi_1(x,\mu^2)
\, , 
& \: &
\displaystyle
\Phi_{Pg}(x,\mu^2)=\frac{f_P^1(\mu_2)}{2\sqrt{2 N_c}} \, \phi_g(x,\mu^2)
\, , 
\end{array}
\label{eq:DAs}
\end{equation}
where the DAs $\phi_8$ and $\phi_1$ are normalized to unity 
\begin{equation}
\int_0^1 dx\, \phi_{i}(x,\mu^2) =1 \, .
\label{eq:qda-norm}
\end{equation}
However, since
\begin{equation}
\int_0^1 dx\, \phi_{g}(x,\mu^2) =0 \, ,
\label{eq:gda-norm}
\end{equation}
there is no such natural way to independently normalize the gluon DA%
\footnote{As we shall explicitly see later on, the DAs satisfy the following
symmetry properties in respect to the longitudinal momentum
fractions $x$:
$\phi_i(x,\mu^2)=\phi_i(1-x,\mu^2)$,
$\phi_g(x,\mu^2)=-\phi_g(1-x,\mu^2)$}. 
Since the
flavor-singlet quark and gluon DAs mix under evolution, it is
convenient to pull out of the gluon DA{} the same factor as for the
flavor-singlet quark one.
In \req{eq:DAs} the particle dependence and the 
mixing behaviour is solely
embedded in the decay constants, while
in a more general approach different distribution amplitudes
$\phi_{P8}$ and $\phi_{P1}$ could be assumed for $P=\eta,\eta'$.
The decay constants are parametrized in a two-angle octet-singlet
mixing scheme 
\cite{Leutwyler97etc,FeldmannKS98}
\begin{eqnarray}
f_{\eta\phantom{'}}^8 = f_8 \cos{\theta_8}\,,
        &\quad& f_{\eta\phantom{'}}^1 = -f_1 \sin{\theta_1}\,, \nn\\
f_{\eta'}^8 = f_8 \sin{\theta_8}\,,
              &\quad& f_{\eta'}^1 =\phantom{-}f_1 \cos{\theta_1}\,.
\label{eq:mix81}
\end{eqnarray}
The numerical values\cite{FeldmannKS98}
$f_8 = 1.26 f_\pi$, $f_1 = 1.17 f_\pi$,
$\theta_8 = -21.2^\circ$, and $\theta_1=-9.2^\circ$ are used in this
work, along with $f_{\pi}=0.131$ GeV.

We note here that alternatively to the octet-singlet basis \req{eq:osbasis}
and the two-angle octet-singlet mixing scheme \req{eq:mix81}, 
the phenomenologically better suited
quark-flavour basis 
($|\qbq\rangle = |(\ubu+\dbd)/\sqrt{2} \rangle$ and 
$|\sbs\rangle$) and quark-flavour mixing scheme \cite{FeldmannKS98} 
were recently suggested.
However, 
since the two-gluon state carries the flavour-singlet quantum numbers
and mixes under evolution with the flavour-singlet component,
octet-singlet basis turns out to be the natural one 
for the hard-scattering leading-twist
analysis which includes the two-gluon components as well.
When the two-gluon states are taken into account,
the DA evolution introduces the appearance of ``opposite''
Fock components in the quark-flavour basis  states%
\footnote{For details, see Sec. III of Ref. \cite{KrollP02}.},
making the calculation unnecessarily difficult and untransparent. 
Furthermore, one has to remember that the one-angle quark-flavour 
mixing scheme has been derived under the assumption that 
the OZI violating effects, and among them
the contributions of the two-gluon components, can be neglected
\cite{Feldmann99}.
Hence, for the calculations involving two-gluon states, one should
use the octet-singlet basis. 
The mixing should be then implemented by the two-angle octet-flavour
mixing scheme whose relation to the quark-flavour scheme
is demonstrated in \cite{FeldmannKS98}.
On the other hand, from the phenomenological
success of the quark-flavour mixing scheme
and the approximate validity of the OZI rule, one should expect that
the effects of the two-gluon components are not excessively large 
in the $\eta$-$\eta'$ system.

The evolution of the octet DA $\phi_8$, being fully analogous
to the pion case, is governed by the evolution equation of the
form
\begin{equation}
  \mu^2 \frac{\partial}{\partial \mu^2}
\phi_8(x,\mu^2)= 
    V(x,u,\alpha_S(\mu^2)) \, \otimes \, 
\phi_8(u,\muF) 
\, ,
\label{eq:evDA8}
\end{equation}
while the singlet and gluon DAs mix under evolution and
the evolution equation takes the matrix form
\begin{equation}
  \mu^2 \frac{\partial}{\partial \mu^2}
\left(
\begin{array}[c]{c}
\phi_1(x,\mu^2) \\[0.2cm]
\phi_g(x,\mu^2) 
\end{array}
\right)=
\left(
\begin{array}{cc}
V_{qq} & V_{qg} \\[0.2cm]
V_{gq} & V_{gg}
\end{array}
\right)(x,u,\alpha_S(\mu^2)) 
\, \otimes \, 
 \left(
\begin{array}[c]{c}
\phi_1(u,\mu^2) \\[0.2cm]
\phi_g(u,\mu^2) 
\end{array}
\right)  
\, .
\label{eq:evDA}
\end{equation}
Here $\otimes$ denotes the usual convolution symbol,
kernels $V$ possess a well defined expansion in
$\alpha_S$ and in this work we are interested only in the
leading-order (LO) evolution%
\footnote{The evolution of the singlet decay constant $f_P^1$
is also to be neglected in that case.}.

The solutions of the LO evolution equation \req{eq:evDA8}
are given in terms of expansion in Gegenbauer polynomials $C_n^{3/2}$ 
\begin{equation}
\phi_{8} (x,\mu^2)  =  6 x (1-x) \left[ 1 
        + 
   \sum_{n=2, 4, \ldots} B_{n}^{\,8}(\mu^2) \; 
          C_{n}^{\,3/2}(2 x -1) \right]\, , 
\label{eq:phi8}
\end{equation}
where the coefficients  $B_{n}^{\,8}(\mu^2)$
evolve according to \cite{sHSA}
\begin{equation}
B_{n}^{\,8}(\mu^2) = B_{n}^{\,8}(\mu_0^2) \, 
               \left( \frac{\als(\mu_0^2)}{\als(\mu^2)} 
               \right)^{\gamma_n^{(0)}/\beta_0} \, ,
\label{eq:Bn8mu}
\end{equation}
$\gamma_n^{(0)}$ are LO anomalous dimensions,
while $B_{n}^{\,8}(\mu_0^2)$ represent nonperturbative
input at the scale $\mu_0^2$.
The LO solutions of \req{eq:evDA} take the more involved form
\begin{eqnarray}
\phi_1(x,\mu^2)  &=& 6 x (1-x) 
\left[ 1 
+ \sum_{n=2,4,\ldots}
{\  B_n^{1}(\mu^2)} \:  C_n^{3/2}(2 x -1) \right]
 \nonumber \\[0.1cm] 
\phi_g(x,\mu^2)  &=&  x^2 (1-x)^2 
 \sum_{n=2,4,\ldots}
 { B_n^{g}(\mu^2)} \:  C_{n-1}^{5/2}(2 x -1)
  \, ,
\label{eq:DAres}
\end{eqnarray}
where
\begin{eqnarray}
{ B_{n}^{\,1}\,(\muF)} & = & B_{n}^{(+)}(\mu_0^2) 
         \left( \frac{\als(\mu_0^2)}{\als(\muF)} \right)^{%
         { \gamma^{\,(+)}_n}/{\beta_0}} 
     + \, 
         { \rho_n^{\,(-)}} \,  B_{n}^{(-)}\,(\mu_0^2) 
         \left( \frac{\als(\mu_0^2)}{\als(\muF)} \right)^{%
         { \gamma^{\,(-)}_n}/{\beta_0}}\hspace*{-1mm} ,
                                         \nn\\[0.2cm]
{ B_{n}^{\,g}\,(\muF)} & = & {\rho_n^{\,(+)}} \, 
               B_{n}^{(+)}\,(\mu_0^2) 
               \left( \frac{\als(\mu_0^2)}{\als(\muF)} \right)^{%
               {\gamma^{\,(+)}_n}/{\beta_0}} 
     +  \, B_{n}^{(-)}\,(\mu_0^2) 
               \left( \frac{\als(\mu_0^2)}{\als(\muF)} \right)^{%
               {\gamma^{\,(-)}_n}/{\beta_0}}
  \, .
\nn \\
\label{eq:Bs}
\end{eqnarray}
Here the coefficients $B_n^{\pm}(\mu_0^2)$, i.e.,
$B_n^{q,(g)}(\mu_0^2)$, represent nonperturbative input at 
scale $\mu_0^2$, while 
$\gamma^{(\pm)}_n$, $\rho_{n}^{(+)}$, $\rho_{n}^{(-)}$,
are defined in terms  of LO anomalous dimensions
(see, for example, \cite{KrollP02}): 
$\gamma^{qq}_n=\gamma_n^{(0)}$,
$\gamma^{gg}_n$, and
\begin{equation}
\gamma^{qg}_n = C_F \; 
              \frac{n (n+3)}{3 (n+1) (n+2)} 
       \, ,
        \qquad
\gamma^{gq}_n = n_f \;  
                     \frac{12}{ (n+1) (n+2)} 
     \, . 
\label{eq:andim}
\end{equation}
Finally, note that, in the limit $\mu^2\to \infty$, 
the octet and singlet DAs evolve into the asymptotic form
$\phi(x)=6 x (1-x)$ and the gluon one to zero.

When calculating the elementary hard-scattering amplitude,
the projection of a collinear $q \bar{q}_i$ state onto a pseudoscalar
meson state is achieved by replacing the quark and antiquark spinors
by 
\begin{equation}
{\cal P}^{i,q}_{\alpha \beta, r s, k l}
= {\cal C}_{i, r s} \frac{\delta_{kl}}{\sqrt{N_c}}
 \left( \frac{\gamma_5 \not \! p}{\sqrt{2}} \right)_{\alpha \beta}
\, ,
\label{eq:qq}
\end{equation}
where $\alpha$ ($r$, $k$) and $\beta$ ($s$, $l$) represent Dirac
(flavour, colour) labels of the quark and antiquark, respectively,
and $p$ denotes the meson momentum ($p^2=0$).
The flavour content is taken into account by the matrices
${\cal C}_8=\lambda_8/\sqrt{2}$ 
and ${\cal C}_1=\mathbf{1}_f/\sqrt{n_f}$,
where $n_f=3$ denotes the number of flavours 
contained in $q\bar{q}_1$.

The projection of a collinear $gg$ state 
onto a pseudoscalar meson state is achieved by replacing 
the gluon polarization vectors 
$\eps^{\mu}(x p, \lambda)$ and $\eps^{\nu}((1-x) p, -\lambda)$
by
\begin{equation}
{\cal P}^{g}_{\mu \nu, a b} =
\frac{i}{2}  \; 
\sqrt{\frac{C_F}{n_f}} \;
\frac{\delta_{ab}}{\sqrt{N_c^2-1}} \;
\eps^{\mu \nu \alpha \beta} \;
\frac{n_{\alpha} p_{\beta}}{n\cdot p} \;
\frac{1}{x(1-x)}
\, ,
\label{eq:gg}
\end{equation}
where $a$, $b$ represent colour indices,
and any vector having the
space components opposite to
$p$ can be taken as $n$ here.
The projector \req{eq:gg} corresponds to
the definition of $\phi_g$, i.e. the anomalous dimensions
$\gamma_n^{qg}$ and $\gamma_n^{gq}$, given by
\req{eq:andim}.

Owing to \req{eq:gda-norm}, there exist freedom in 
defining the gluon DA.
Suppose we change $\phi_{g}$ by a
factor $\sigma$
\begin{equation}
    \phi^{\,\sigma}_{Pg} = \sigma\, \phi_{Pg}\,.
\label{eq:phigsigma}
\end{equation}
Inspection of Eq. \req{eq:evDA},
(or equivalently of Eqs.\ (\ref{eq:DAres}-\ref{eq:Bs}) )
reveals the following.
Since the singlet and gluon
DA are connected by evolution and
in order to leave the quark DA $\phi_1$ unchanged,
the change of the definition of the gluon
DA \req{eq:phigsigma}
has to be converted into a change 
of $V_{qg}$ and $V_{gq}$, or equivalently 
into a change of the off-diagonal anomalous
dimensions $\gamma_{qg}$ and $\gamma_{gq}$
and the  $B_{Pn}^{(-)}$. 
Hence, \req{eq:phigsigma} is equivalent to 
\begin{equation}
\gamma^{\,qg, \sigma}_n = \frac{1}{\sigma} \, \gamma^{\,qg}_n\,,
 \qquad
\gamma^{\,gq, \sigma}_n = \sigma \, \gamma^{\,gq}_n\,,
\label{eq:gammasigma}
\end{equation}
and
$
{B}^{(-)\,\sigma}_{Pn}\,(\mu_0^2)=\sigma B^{(-)}_{Pn}\,(\mu_0^2)
$,
which then implies
$B^{\,g\,\sigma}_{Pn}\,(\mu^2)=\sigma\, B^{\,g}_{Pn}\,(\mu^2)$
and $B^{\,q\,\sigma}_{Pn}\,(\mu^2)= B^{\,q}_{Pn}\,(\mu^2)$.
On the other hand, 
since any physical quantity 
 must be independent of the choice of the convention,
any change of the definition of
the gluon DA is naturally 
accompanied by a corresponding change in the elementary hard-scattering
amplitude.
Namely,
the projection \req{eq:gg} of the $gg$ state onto a pseudoscalar meson
state is to be modified by a factor $1/\sigma$, i.e.
\begin{equation}
{\cal P}^{g\,\sigma}_{\,\mu \nu} =
   \frac{1}{\sigma}\, {\cal P}^{g}_{ \, \mu \nu}
  \,,
\label{eq:ggsigma}
\end{equation}
and the elementary hard-scattering amplitude becomes altered accordingly.
In the literature
\cite{Terentev81,ShifmanV81etc,BelitskyM98etc} 
one encounters various conventions for $\gamma_n^{qg}$ and $\gamma_n^{gq}$,
but the corresponding definition of the gluon projector
${\cal P}^g$ was often omitted, and it is crucial that these
two ingredients of the leading-twist calculation are consistently defined.
In Ref. \cite{KrollP02} a consistent set of conventions
\req{eq:andim} and \req{eq:gg} was fixed
and tested on the NLO calculation of the $\eta$, $\eta'$-photon
transition form factor. 
The relations \req{eq:gammasigma} and \req{eq:ggsigma}
then enable us to 
make a connection with other conventions
(note that the input coefficients $B_n^{(-)}(\mu_0^2)$,
i.e. $B_n^{g}(\mu_0^2)$, are also convention dependent).

\section{{\it{\Large Applications}}}

\begin{figure}[th]
\centerline{\includegraphics[height=7.5cm]{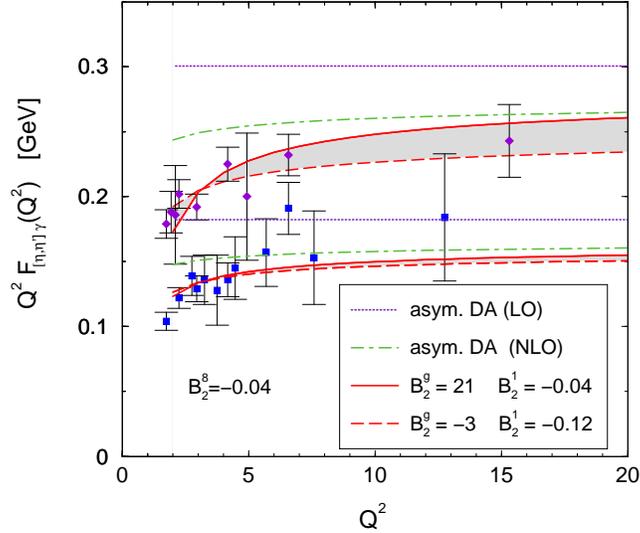}}
\caption{$\eta$ (below) and $\eta'$ (above) transition form factors. 
The shaded area corresponds to 
the range of $B_2^1(\muO)$ and $B_2^g(\muO)$
given in Eq. \protect\req{eq:fit81g}.}
\label{f:etff}
\end{figure}

First, we turn to the NLO calculation of the 
$\eta$, $\eta'$--photon transition form factor,
i.e. to the evaluation of the $\gamma^* \gamma \to \eta (\eta')$
hard-scattering amplitude.
The form factor can be expressed as a sum
\begin{equation}
F_{P\gamma}=F_{P\gamma}^8(Q^2) \, + \, 
            F_{P\gamma}^{1g}(Q^2)
\, ,
\end{equation}
where $Q^2$ represents the photon virtuality.
The flavour-octet contribution $F_{P\gamma}^8(Q^2)$
can be obtained from the pion--photon transition form factor result
(see \cite{MelicNP01} and references therein);
one only has to take into account the proper flavour factor.
The contributions of the flavour-singlet and two-gluon
components contained in 
\begin{equation}
    F_{P\gamma}^{\,1g}(Q^{2})  =  
\left( 
T_{H,1}(x,Q^{2},\muF) \quad
 T_{H,g}(x,Q^{2},\muF)  \right)
 \, \otimes \,
 \left( 
\begin{array}[c]{c}
 \Phi_{P1}(x,\muF) \\[0.2cm]
 \Phi_{Pg}(x,\muF)  
\end{array}
 \right) 
\, ,
\end{equation}
were calculated
in Ref. \cite{KrollP02}. 
Following the recent analysis of the pion--photon 
transition form factor\cite{MelicNP01},
a detailed NLO analysis was performed
taking into account both the hard-scattering 
part and the perturbatively calculable DA part. 
The cancellation of the collinear singularities
present in the parton-subprocess amplitude with the ultraviolet (UV)
singularities appearing in the unrenormalized DAs%
\footnote{Note that 
the renormalization introduces mixing of the composite operators
$\bar{\Psi}(-z) \, \gamma^+ \gamma_5 \, \Omega \, \Psi(z)$
and $G^{+ \alpha}(-z) \, \Omega \, \widetilde{G}^{\: \: \: +}_{\alpha}(z)$
in terms of which the quark singlet and gluon DAs are
defined, respectively.}
offered the most crucial test of the consistency of our
set of conventions for singlet and gluon DAs and projectors.
Using the mixing scheme defined in Eq. \req{eq:mix81},
the NLO leading-twist prediction for the $\eta$ and $\eta'$ 
transition form factors was obtained.
Owing to quality and quantity of the experimental data\cite{exptff},
the Gegenbauer series \req{eq:Bn8mu} and \req{eq:DAres}
were truncated at $n=2$, and the results were then fitted to the data.
For $Q^2\geq 2$ GeV$^2$ and $\mu_0=1$ GeV, the results of the fits read 
\begin{equation}
B_2^8(\muO)  = -0.04 \pm 0.04 \quad 
B_2^1(\muO) = -0.08 \pm 0.04  \quad
B_2^g(\muO)  = 9 \pm 12
\, .
\label{eq:fit81g}
\end{equation}
The existing experimental data and their quality allow us
to obtain not more than a constraint on the value of $B_2^g$.
As expected, we have observed  a strong correlation between $B_2^1$ and $B_2^g$.
The quality of the fit as well as the sensitivity
of the results on the size of two-gluon components%
\footnote{Since $B_2^1$ and $B_2^g$ are correlated, the shaded area in
Fig. \ref{f:etff} corresponds to the change of both of these coefficients.
Nevertheless, the variation of $B_2^g$ is numerically dominant.}
can be seen from Fig. \ref{f:etff}.  

As a next application,
the $\eta'$--gluon transition form factor, 
i.e. $g^* g^* \eta'$ vertex,
turns to be a natural choice.
In contrast to the $\eta$, $\eta'$--photon transition
form factor, the two-gluon components contribute 
to $g^* g^* \to \eta (\eta')$ already at LO in $\als$ 
and the contribution of the
$q \bar{q}_8$ component vanishes.
A consequence of the latter is that the $\eta$--gluon
transition form factor is much smaller
than the $\eta'$ one.
A reliable determination of the $g^* g^* \eta'$ vertex is of
importance for the calculation of a number of decay processes such as 
$B \to \eta' X$ \cite{AtwoodS97}, $B \to \eta' K$,
or of the hadronic production process $p p  \to \eta' X$. 
To leading-twist order the $g^* g^* \eta'$ vertex has been first
calculated in Refs. \cite{MutaY99etc}.
In \cite{KrollP02} it was reanalyzed
using our set of conventions, the previous calculations were examined
and corrected, and the numerical predictions using
the Gegenbauer coefficients \req{eq:fit81g} were provided.
As expected, it was shown that the $g^* g^* \eta'$ vertex
is quite sensitive to the two-gluon components.

\begin{figure}[h]
\centerline{
\includegraphics[height=5.5cm]{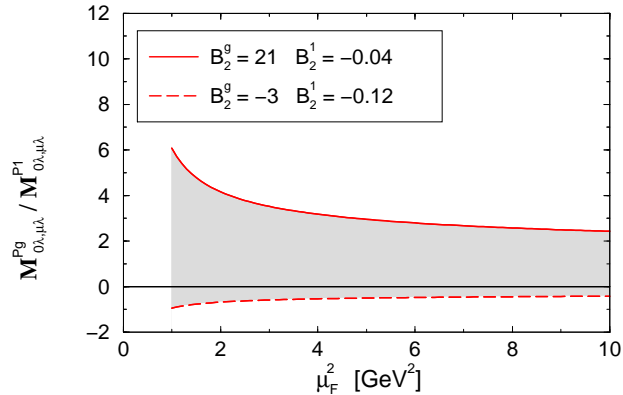}
}
\caption{Ratio of the gluon and singlet quark amplitudes
for the photoproduction of $\eta'(\eta)$ mesons as a function
of $\mu_F^2$. 
The shaded area corresponds to 
the range of $B_2^1(\muO)$ and $B_2^g(\muO)$
given in Eq. \protect\req{eq:fit81g}.}
\label{f:ratio}
\end{figure}

In Ref. \cite{KrollP02} our formalism was applied also to
the deeply-virtual and wide-angle electroproduction of 
$\eta$ and $\eta'$ mesons with longitudinal photons.
It was found that in the former the two-gluon contributions
were suppressed, while in the latter they
could be important depending on the size of the $B_2^g$ coefficient.
Here we have extended this analysis to the photoproduction of 
$\eta$ and $\eta'$ mesons
calculated in the handbag approach in which  
the process $\gamma p \to \eta (\eta') p$
factorizes in the subprocess amplitude 
$\gamma q \to \eta (\eta') q$ and
soft proton matrix elements.
The meson is again generated by the leading-twist
mechanism.
As in the case of the wide-angle electroproduction,
the two-gluon contributions could be substantial 
and we illustrate that by displaying 
the ratio of the $gg$ and $q \bar{q}_1$ contributions
(see Fig. \ref{f:ratio}).

Finally, we mention some further applications
that can be found in recent literature:
in Ref. \cite{BenekeN02} 
the previously explained formalism 
was applied to the $B\to \eta'K^{(*)}$ process,
the process $\Upsilon(1S)\to \eta'X$ was analyzed in \cite{AliP03} 
obtaining further restrictions on the $B_2^g$ coefficient,
while some modifications of the leading-twist formalism
were introduced in \cite{newgg}.

\section{{\it{\Large Description of glueballs}}}

Last but not least, we would like to comment on
the possible application of this formalism
to the description of glueballs.

The pioneering work was done in Ref. \cite{WakelyC92},
were the pseudoscalar glueballs were described in the
standard hard-scattering approach while the gluon DA was
obtained from the QCD sum rules.
The results were then applied to the $\gamma \gamma \to G \pi$
($G=$glueball) process, but the mixing with the $q \bar{q}_1$
state and the evolution were neglected.
However, in a consistent approach, the mixing of $gg$ and
$q \bar{q}_1$ components under evolution should 
be taken into account.

Let us examine from the purely theoretical point of view
the possible description of the glueball states 
in leading-twist formalism.
In the pseudoscalar case, one should describe the glueball
using the $\Phi_{Pg}$ and $\Phi_{P1}$ DAs of the form given
by Eqs. \req{eq:DAs} and \req{eq:DAres},
where $P$ now denotes the pseudoscalar glueball.
The decay constant $f_{P1}$ as well as the $B_{n}^g$ and $B_{n}^1$
coefficients are unknown. 
For simplicity reasons, let us again take only
$n=2$ and then compare $\phi_1$ and $\phi_g$ from \req{eq:DAres}. 
One can easily see that in order to 
have the dominantly glueball state, $B_2^g$ should be much larger than $1$,
i.e., than the  ``leading'' term in the expansion of $\phi_1$. 
This is a condition which may not be trivially satisfied,
especially since $B_2^g$ decreases with $\mu^2$ and,
in the limit $\mu^2 \to \infty$,  the pseudoscalar gluon DA vanishes,
leaving us only with the $q \bar{q}_1$ contribution.

The situation looks more favourable in the scalar case
where we describe glueballs in terms of $\Phi_{Sg}$ and $\Phi_{S1}$,
$S$ being the scalar glueball state.
The equivalent of Eq. \req{eq:DAres} is
given for the scalar case by
\begin{eqnarray}
\phi_{S1}(x,\mu^2)  &=&  x (1-x) 
+ \sum_{n=2,4,\ldots}
{\  B_{Sn}^{1}(\mu^2)} \:  C_{n-1}^{3/2}(2 x -1) 
  \, ,
 \nonumber \\[0.1cm] 
\phi_{Sg}(x,\mu^2)  &=&  30 x^2 (1-x)^2 
\left[ 1 +
 \sum_{n=2,4,\ldots}
 { B_{Sn}^{g}(\mu^2)} \:  C_{n}^{5/2}(2 x -1)
\right]
  \, .
 \nonumber \\[0.1cm] 
\label{eq:DAresS}
\end{eqnarray}
One can see that, in a sense, the role of the gluon and quark singlet
DAs is here reversed.
The gluon DA is now symmetric and well normalized
(compare $\int_0^1 dx \phi_{Sg}=1$ and $\int_0^1 dx \phi_{S1}=0$
with Eqs. (\ref{eq:qda-norm}-\ref{eq:gda-norm})), 
and in the limit of $\mu^2\to \infty$, 
the quark singlet DA vanishes, while the gluon one takes the asymptotic form
$\phi_g(x)=30 x^2 (1-x)^2$.
In order to have the dominantly glueball state,  it is now sufficient
that $B_{S2}^1$ is sufficiently smaller than $1$
and this may be expected, especially since
$B_{S2}^1$ decreases with $\mu^2$.

The analysis of both the pseudoscalar and scalar glueballs 
along these lines is underway.

\section{{\it{\Large Conclusions}}}

In this work we have reviewed the leading-twist 
hard-scattering formalism for the description of $\eta$
and $\eta'$ mesons with two-gluon components included.
The theoretical and numerical results of Ref. \cite{KrollP02}
have been summarized and applied further to the photoproduction
of $\eta$ and $\eta'$ mesons as well as to the possible description
of pseudoscalar and scalar glueballs.
The processes such as $g^* g^* \to \eta'$, wide-angle electroproduction
and photoproduction
of $\eta$ and $\eta'$ mesons show sensitivity to two-gluon
contributions.
Future data should allow to pin down the gluon DA, 
while the description of glueballs offers a new
interesting area of application of this formalism.


\begin{center}
{\em Acknowledgments}
\end{center}
  I would like to thank P. Kroll for fruitful collaboration and
  lengthy discussions.
  This work was supported by the Ministry of Science and Technology
  of the Republic of Croatia under Contract No. 0098002.


%
%
%
%
%
%
%

\end{document}